# A Glass-Box Deep-Learning Method for Electrical Energy System Modeling Based on Kolmogorov-Arnold Network

Zhenghao Zhou, *Student Member, IEEE*, Yiyan Li, *Member, IEEE*, Zelin Guo, *Student Member, IEEE*, Zheng Yan, *Senior Member, IEEE*, and Mo-Yuen Chow, *Fellow, IEEE*

***Abstract*—Deep learning methods have been widely used as an end-to-end modeling strategy of electrical energy systems because of their conveniency and powerful pattern recognition capability. However, due to the "closed-box" nature, deep learning methods have long been blamed for their poor interpretability when modeling a physical system. In this paper, we introduce a novel neural network structure, Kolmogorov-Arnold Network (KAN), to achieve "glass-box" modeling for electrical energy systems to enhance the interpretability. The most distinct feature of KAN lies in the learnable activation function together with the sparse training and symbolification process. Consequently, KAN can express the physical process with concise and explicit mathematical formulas while remaining the nonlinear-fitting capability of deep neural networks. Simulation results based on three electrical energy systems demonstrate the effectiveness of KAN in the aspects of interpretability, accuracy, robustness and generalization ability.**

***Index Terms*—Kolmogorov-Arnold network, electrical energy system modeling, interpretable neural network, learnable activation function.**

## I. Introduction

WITH the increasing integration of renewable energy and flexible loads, the power systems are equipped with large amounts of power electronic devices and are becoming increasingly complex [1][2]. Accurate modeling for the electronic power systems is critical to understanding the new system dynamics and to securing the system operation. Physics-based methods have been widely implemented to model the electrical systems. Following the First-Principles Theory, physics-based methods try to describe the internal physical process of the electrical systems by using explicit mathematical equations, which is interpretable and usually has stable performance. For instance, researchers have developed a series of Equivalent Circuit Models (ECM) to characterize the lithium-ion battery systems [3][4]. However, due to the complex dynamics and nonlinearity of the electrical systems, the accuracy of a static physical model may fade with time. Moreover, developing accurate physical models for the large amounts of electrical systems integrated to the grid is labor-intensive and requires heavy domain knowledge, which is difficult to accomplish.

To further enhance the modeling accuracy and efficiency, data-driven methods are gaining more attention in recent years with the development of Artificial Intelligence (AI) and data analytics. Data-driven methods can learn the patterns of the electrical systems in an end-to-end fashion from the field measurements. As a result, data-driven methods require little domain knowledge and can update automatically based on the latest data samples. For example, Tian et al. propose a data-driven modeling approach for virtual synchronous generators based on Long Short-Term Memory networks (LSTM) [5]. Liao et al. apply neural network to the frequency-domain impedance modeling of power electronic systems [6]. Zheng et al. introduce a novel equivalent model based on LSTM and Recurrent Neural Network (RNN) to accurately represent active distribution networks in modern power systems [7]. However, as an end-to-end modeling strategy, data-driven approaches have been criticized for their limited interpretability, especially deep learning-based models. These models can enhance the accuracy of modeling complex systems by increasing the number of parameters, but this comes at the cost of reduced model interpretability [8]. The trade-off between model accuracy and model interpretability limits the application of machine learning [9]. Such "closed box" feature impairs the application value of the data-driven models because system operators can hardly put trust on the results that are not interpretable.

In general, interpretability denotes the degree to which humans are able to comprehend the mechanisms and reasoning processes behind a model's predictions. In the modeling of complex physical systems, physics-based

This work was supported by National Natural Science Foundation of China under Grant 52307121, and also supported by Shanghai Sailing Program under Grant 23YF1419000. (Corresponding author: Yiyan Li.)

Zhenghao Zhou, Yiyan Li, Zelin Guo are with the College of Smart Energy, Shanghai Non-Carbon Energy Conversion and Utilization Institute, and Key Laboratory of Control of Power Transmission and Conversion, Ministry of Education, Shanghai Jiao Tong University, Shanghai, 200240, China. (e-mail: zhenghao.zhou@sjtu.edu.cn, yiyan.li@sjtu.edu.cn, gzl1996@sjtu.edu.cn).

Zheng Yan is with the Key Laboratory of Control of Power Transmission and Conversion, Ministry of Education, and the Shanghai Non-Carbon Energy Conversion and Utilization Institute, Shanghai Jiao Tong University, Shanghai 200240, China. (e-mail: yanz@sjtu.edu.cn)

Mo-Yuen Chow is with the University of Michigan - Shanghai Jiao Tong University Joint Institute, Shanghai Jiao Tong University, Shanghai, 200240, China. (email: moyuen.chow@sjtu.edu.cn)

approaches are characterized by high interpretability but limited predictive accuracy due to their reliance on simplified governing equations, whereas machine learning-based methods demonstrate superior accuracy in capturing nonlinear dynamics but suffer from a lack of interpretability owing to their data-driven "closed-box" nature [10], [11], [12].

To address the “closed box” issue of the deep learning models, researchers have tried different methodologies to improve the model interpretability. *First*, attention mechanism is introduced to the neural network models [13]. Attention is a specially-designed neural network module that can achieve dynamic weighting for the input vectors. Consequently, researchers can understand which features the neural network is focusing on by visualizing or analyzing the attention module outputs, thus improving the interpretability of the neural network decision process [14], [15]. *Second*, Physics-Informed Neural Network (PINN) have been proposed to enhance the neural network interpretability by integrating physical laws into the networks [16]. This integration can be accomplished via four main paradigms: modification to the loss function, specialized initialization technique, tailored structural design, and the hybrid physics-data-driven model. For instance, in [17], the loss function are constrained by physical laws to enhance the model stability and interpretability. In [18], the connection between the latent layers and the final model outputs is established based on the converter dynamic equations, while the loss function is composed of the dynamic variable errors. In [19], the hybrid learning is leveraged to implement the dynamic state estimation of synchronous generators. *Third*, sensitivity analysis tools, such as SHapley Additive exPlanations (SHAP) [20] and Local Interpretable Model-agnostic Explanations (LIME) [21], are introduced to quantify the influence of the input variables on the final model outputs, which contribute to the neural network interpretability. However, due to the large number of model parameters and the hyper-complex internal calculation process, the “closed box” nature of deep learning models has not been fundamentally changed.

In this paper, we introduce the Kolmogorov-Arnold Network (KAN) [22], a novel neural network structure that can achieve “glass-box” modeling for physical systems. The most distinct feature of KAN is that KAN replaces the fixed activation functions at the nodes of traditional neural networks with trainable activation functions at the edges, shown as Fig. 1. Like traditional neural networks, KAN can be initialized with complex network structure to guarantee the learning ability. However, only a few edges will be remained after the sparse training process, which will be further replaced by common basis functions (e.g., quadratic function, sinusoidal function, linear function, etc.). As a result, KAN can express the physical process with concise and explicit mathematical formulas instead of large-scale neuron computation, which significantly improves the interpretability while remaining the conveniency and accuracy as a deep learning method. The main contributions of our paper are considered threefold:

1) We introduce the KAN with the objective of deriving symbolic representations that support transparent "glass-box" modeling in electrical energy systems.

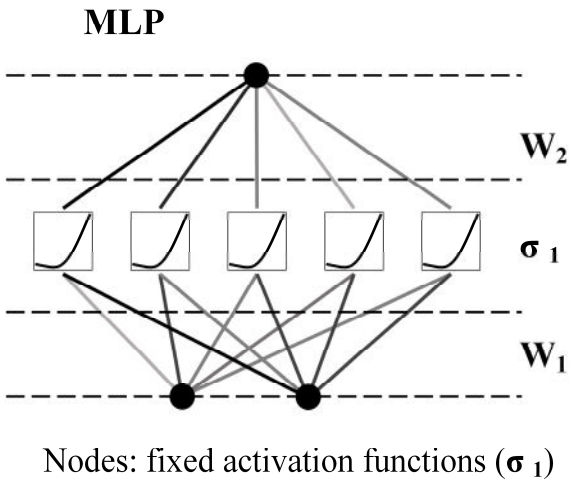

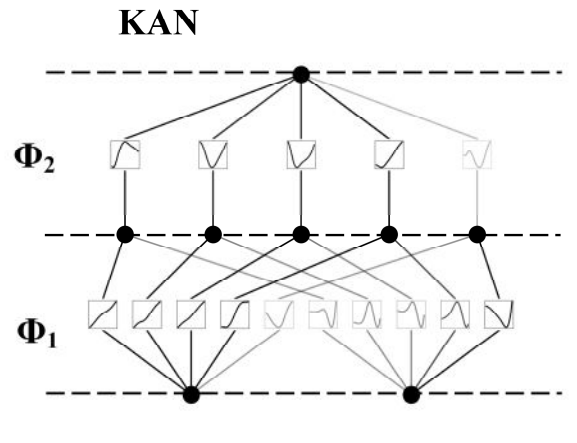

Fig. 1. Network structure of MLP and KAN, with darker colors in MLP representing larger weights.

2) We purpose a novel neural network-based function classifier to improve the efficiency and accuracy of the symbolification process in the vanilla KAN model.

3) We comprehensively evaluate the performance of KAN in the aspects of interpretability, accuracy, robustness and generalization ability when modeling electrical energy systems.

To the best of the author's knowledge, this paper is the first attempt to implementing KAN to solve the modeling problem of physical systems.

The rest of the paper are organized as follows: Section II introduces the theoretical foundation, the network structure and the application methodology of KAN. Section III provides simulation results of KAN in modeling different electrical energy systems. Section IV concludes this paper.

## II. Methodology

In this section, we first introduce the theoretical basis of KAN: the Kolmogorov-Arnold representation theorem. Then we introduce the structure of the KAN network. Next, we introduce the training process of KAN, and analyze the interpretability of KAN. Finally, we explain the application of KAN in solving unsupervised learning tasks.

### *A. Kolmogorov-Arnold Representation Theorem*

The Kolmogorov-Arnold representation theorem, proposed by Andrey Kolmogorov and Wolfgang Arnold, reveals that any smooth multivariable continuous function on a bounded domain can be precisely represented by a finite polynomial combination of a series of univariate continuous functions [23].

For any smooth function $f$: $[0,1]^n \rightarrow \mathbb{R}$ defined on a bounded domain, there exists a series of continuous univariate functions $\varphi$ and $\psi$, such that:

$$f(x)=\sum_{q=1}^{2n+1}\psi_q\left(\sum_{p=1}^{n}\varphi_{p,q}(x_p)\right), \tag{1}$$

where $x_p \in (x_1,\ldots,x_n)$, $\varphi_{p,q}$: $[0,1]^n \rightarrow \mathbb{R}$ and $\psi_q$: $\mathbb{R} \rightarrow \mathbb{R}$ are continuous functions. Based on this theorem, learning a high-dimensional function can be reduced to learning a polynomial combination of univariate functions, which provides the theoretical foundation for establishing the KAN network.

### *B. Kolmogorov-Arnold Network*

Based on the Kolmogorov-Arnold representation theorem, KAN is designed by Liu et al. with learnable univariate activation functions, as shown in Fig. 1. Because all activation functions are set up on the network edges, KAN can interpret each activation function $\varphi$ as the basis functions $b(x)$ plus the spline function $spline(x)$

$$\varphi = \omega_b b(x) + \omega_s spline(x), \tag{2}$$

$$b(x) = silu(x) = \frac{x}{1+e^{-x}}, \tag{3}$$

where $\omega_b$ and $\omega_s$ are the weights. $spline(x)$ is further expressed as a linear combination of B-spline curves $B_i(x)$ with learnable coefficients $c_i$

$$spline(x) = \sum_i c_i B_i(x), \quad (i = G + k) \tag{4}$$

where $G$ is the number of grid intervals, and $k$ is the piecewise polynomial order of splines. Thus, a KAN layer with $n_{in}$ inputs and $n_{out}$ outputs can be defined as a set of univariate activation functions:

$$\mathbf{\Phi} = \{\varphi_{q,p}\}, \quad p = 1,2,\cdots,n_{in}, \quad q = 1,2,\cdots,n_{out} \tag{5}$$

where the functions $\varphi_{q,p}$ have trainable parameters. According to (1), to approximate a function, the network can be realized as a simple combination of two KAN layers. The first layer has an input of $n_{in}$=$n$ and an output of $n_{out}$=2$n$+1, while the second layer has an input of $n_{in}$=2$n$+1 and an output of $n_{out}$=1. However, in practice, KAN is not subject to this constraint and more KAN layers can be stacked. A $N$-layer KAN network can be represented by an integer array

$$[n_0, n_1, ..., n_l, n_{l+1}, ..., n_N], \tag{6}$$

where $n_l$ is the number of nodes in the $l^{th}$ layer of KAN model. When calculating the KAN outputs of the $(l+1)^{th}$ layer, it can be represented by matrix operation as follows:

$$\mathbf{x}_{l+1} = \underbrace{\begin{pmatrix} \varphi_l^{(1,1)}(\cdot) & \varphi_l^{(1,2)}(\cdot) & \cdots & \varphi_l^{(1,n_l)}(\cdot) \\ \varphi_l^{(2,1)}(\cdot) & \varphi_l^{(2,2)}(\cdot) & \cdots & \varphi_l^{(2,n_l)}(\cdot) \\ \vdots & \vdots & \ddots & \vdots \\ \varphi_l^{(n_{l+1},1)}(\cdot) & \varphi_l^{(n_{l+1},2)}(\cdot) & \cdots & \varphi_l^{(n_{l+1},n_l)}(\cdot) \end{pmatrix}}_{\mathbf{\Phi}_l} \mathbf{x}_l, \tag{7}$$

where $\varphi_l^{(i,j)}$ refers to the activation function connecting the $i^{th}$ neuron in layer $l$ and the $j^{th}$ neuron in layer $l$+1. Using $\mathbf{\Phi}_l$ to represent the function set in the $l^{th}$ KAN layer, then a KAN model with $L$ layers can be represented as

$$\mathrm{KAN}(\mathbf{x}) = (\mathbf{\Phi}_{L-1} \circ \mathbf{\Phi}_{L-2} \circ \cdots \circ \mathbf{\Phi}_1 \circ \mathbf{\Phi}_0)\mathbf{x}. \tag{8}$$

As a comparison, the vanilla MLP network places the fixed activation function σ on nodes, and the learnable weight $\mathbf{W}$ on edges, as shown in Fig. 1. An $L$-layer MLP can be written as

$$\mathrm{MLP}(\mathbf{x}) = (\mathbf{W}_{L-1} \circ \sigma \circ \mathbf{W}_{L-2} \circ \sigma \circ \cdots \circ \sigma \circ \mathbf{W}_0)\mathbf{x}. \tag{9}$$

### *C. Training Process and Interpretability Analysis of KAN*

The training process of KAN is summarized in Fig. 2, including four stages: initialization, sparsification, symbolification and parameter refining.

1) *Initialization*. Similar to traditional neural networks, KAN is initialized by setting up the network structure, the activation functions and the trainable parameters. According to equation (1), a KAN network should include at least 2 layers to guarantee the function approximation capability. In this section, we select a 2-layer KAN network with the number of nodes [2,5,1] as an example to demonstrate the KAN training process.

2) *Sparsification*. After initialization, the training process of KAN focuses on sparsifying the initial network configuration to reduce the network complexity and improve the interpretability. To do this, regularization and pruning strategies are implemented in the training process.

*First*, regularization is achieved by introducing both L1 regularization term and entropy regularization term to the KAN loss function. For an activation function $\varphi$, the L1 norm is defined as the averaged value of the function outputs over all $N_s$ training samples

$$|\varphi|_1 = \frac{1}{N_s}\sum_{k=1}^{N_s}|\varphi(x_k)|. \tag{10}$$

Then for a KAN layer $\mathbf{\Phi}$ with $n_{in}$ inputs and $n_{out}$ outputs, the L1 regularization term can be calculated by summing up the L1 norm over all activation functions included in $\mathbf{\Phi}$

$$|\mathbf{\Phi}|_1 = \sum_{i=1}^{n_{in}}\sum_{j=1}^{n_{out}}|\varphi_{i,j}(x_i)|_1. \tag{11}$$

By minimizing the L1 regularization term in the loss function, the training process punishes complex network structures that leads to large output values, driving the network to become sparser and simpler.

Entropy regularization further ensures the KAN sparsity. The entropy of a KAN layer $\mathbf{\Phi}$ measures the distribution uniformity of all activation functions included in this layer, which is defined as

$$S(\mathbf{\Phi}) = -\sum_{i=1}^{n_{in}}\sum_{j=1}^{n_{out}}\frac{|\varphi_{i,j}|_1}{|\mathbf{\Phi}|_1}\log\left(\frac{|\varphi_{i,j}|_1}{|\mathbf{\Phi}|_1}\right). \tag{12}$$

By minimizing the entropy, the training process encourages more uniformly distributed activation functions in $\mathbf{\Phi}$ to reduce the number of activation functions being activated simultaneously. In other words, there is a "dropout" rate among activation functions to further sparsify the KAN network.

The total loss function $\mathcal{L}_{total}$ is defined as the network prediction loss $\mathcal{L}_{pred}$ plus the two regularization terms

$$\mathcal{L}_{total} = \mathcal{L}_{pred} + \lambda\left(\mu_1\sum_{l=0}^{L-1}|\mathbf{\Phi}_l|_1 + \mu_2\sum_{l=0}^{L-1}S(\mathbf{\Phi}_l)\right), \tag{13}$$

where $\mu_1$, $\mu_2$ and $\lambda$ are weighting factors. Typically, $\mu_1 = \mu_2 = 1$. The $\mathcal{L}_{pred}$ is defaulted as mean squared error.

Note that in the training process, KAN employs an adaptive grid refinement mechanism to better align the model's internal representation with the evolving data distribution. Initially, the grid is uniformly initialized based on the model weights. However, as training progresses and activation distributions shift, this initial setup may become suboptimal. To address this issue, KAN dynamically adjusts the grid by first sorting the input activations of the current layer. Then KAN selects boundary points at equal quantile intervals to construct an adaptive grid, which reflects the empirical distribution of the data. Meanwhile, a uniform grid spanning the full activation range—augmented with a small margin (0.01) on both ends—is generated to ensure robustness. These two grids are linearly interpolated using a blending hyperparameter, where extreme values select one grid exclusively, and intermediate values yield a mixture. After each grid update, the spline coefficients are recomputed, preserving the function output while maintaining continuity. This adaptive strategy enables finer resolution in high-density regions and coarser approximations in sparse ones.

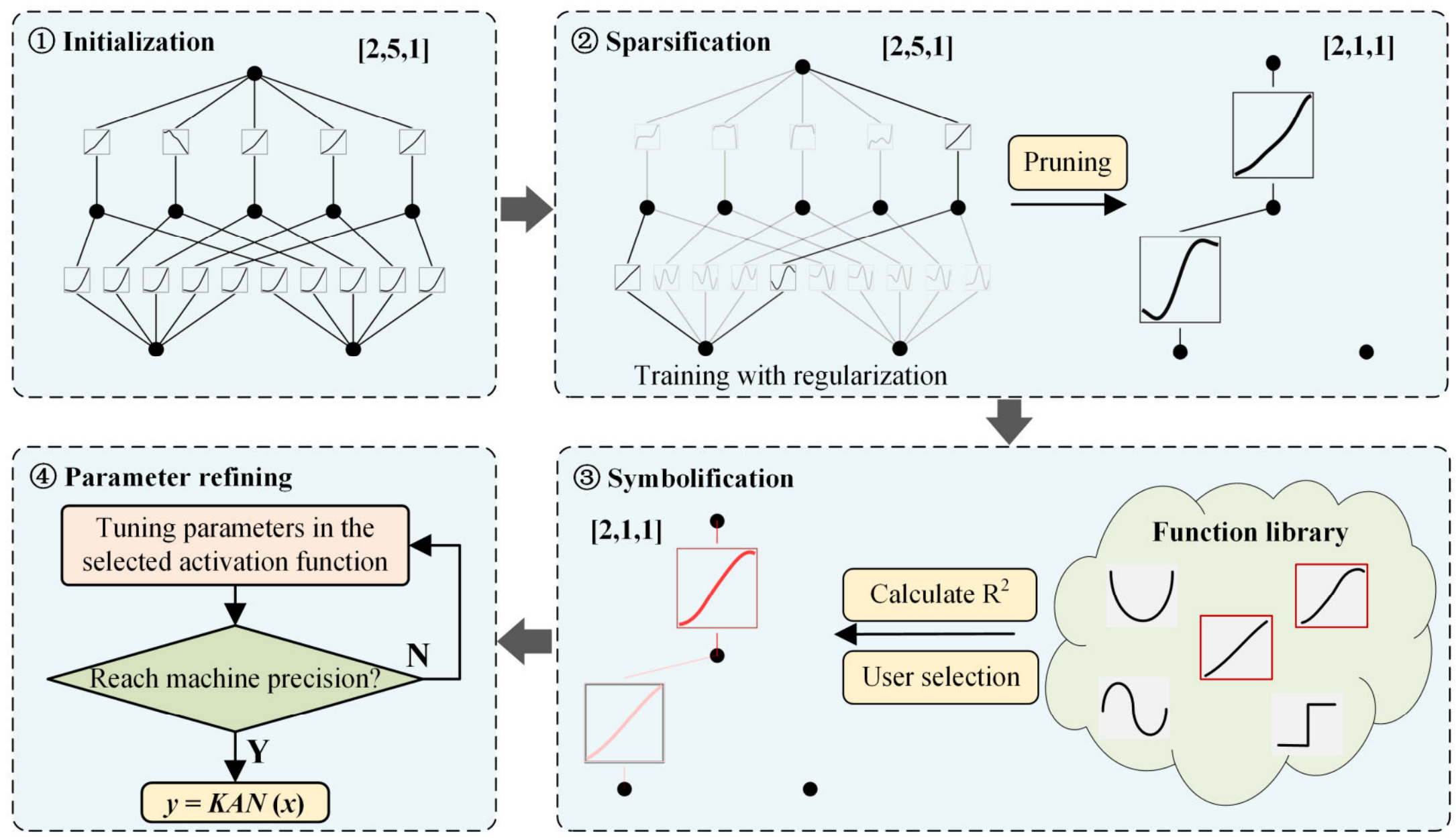

Fig. 2. The training process of KAN for symbolic regression. Higher transparency of the activation function plots refers to smaller activation function magnitude.

*Second*, after the KAN network is trained with regularization losses, the trained network will be further pruned at the node level to enhance the sparsity. For each node, specifically the $i^{th}$ neuron within the $l^{th}$ layer, the input score and the output score are defined as follows:

$$I_{l,i} = \max_{k}(|\varphi_{l-1,k,i}|_1), \quad O_{l,i} = \max_{j}(|\varphi_{l+1,j,i}|_1). \tag{14}$$

Since a neuron has more than one edges connected to it, formula (14) calculates the magnitude of all edges and takes the highest value as the final score. If the input and output scores of a node exceed a given threshold, this node is considered important and will be remained. All other nodes are considered unimportant and will be pruned together with their connected edges.

3) *Symbolification*. After sparsification, the KAN network will be simplified with a limited number of nodes and edges remained. In this stage, the activation functions on the network edges, which have been expressed by a combination of spline functions shown in equation (2), will be replaced by similar common basis functions to enhance the interpretability. Common basis functions refer to naïve single-variable functions, such as quadratic function, sinusoidal function, linear function, etc. This step can either be user-defined or auto-selected. Users can set up the replacement based on the activation function plots. Otherwise, the model will automatically search the best replacement in the predefined function library by calculating $R^2$ correlation scores. As the common functions usually have simpler symbolic expressions and clearer physical meanings, the interpretability of KAN is further improved.

Note that in the symbolification process, the size of the function library significantly impacts the computational cost. Because the activation function must compute $R^2$ correlation scores with every candidate function exhaustively, the total computation time increases linearly with the size of the library.

4) *Parameter refining*. After all activation functions are finalized, the training process starts to fine-tune the parameters in the activation functions to improve the model performance. The training process will end when the loss drops to machine precision.

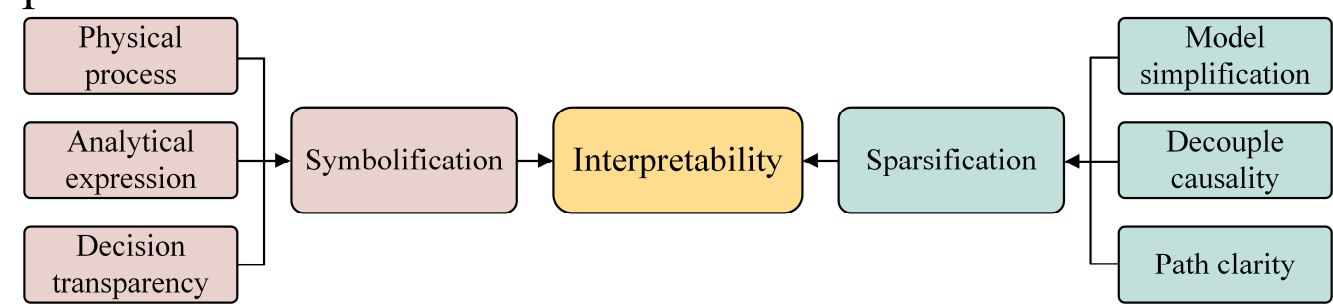

Fig. 3. The interpretability analysis of KAN.

Based on the above training process, the interpretability of KAN mainly lies in two aspects, as shown in Fig. 3. First, the trained KAN network is highly sparse with only a few nodes and edges remained. The simplified model easily decouples the causal relationships between inputs and outputs, providing a clear pathway from input to output. As a result, the KAN network can be expressed by simple and explicit symbolic equations. As a comparison, although traditional neural networks can also be re-written as symbolic equations shown as equation (9), the expression will be too complicated to interpret because of large amounts of parameters and complex neuron connections. Second, because of the symbolification process, the trained KAN network is essentially a combination of common functions, which are easier to be interpreted along with the actual physical process of an electrical energy system. The final analytical expressions, composed of common functions, provide a transparent exposition of the model's decision-making process.

## D. *Unsupervised Application of KAN*

The modeling process of an electrical energy system is usually a supervised learning process with specified model inputs and outputs. However, KAN can be also implemented in an unsupervised learning fashion to uncover the underlying relationship among variables when the input-output mapping is not clear. Assume there are $d$ variables ($x_1$, $x_2$,…, $x_d$) with unknown dependency. Then the machine learning problem can be formulated as identifying a non-zero function $f$:

$$f(x_1, x_2, ..., x_d) \approx 0. \tag{15}$$

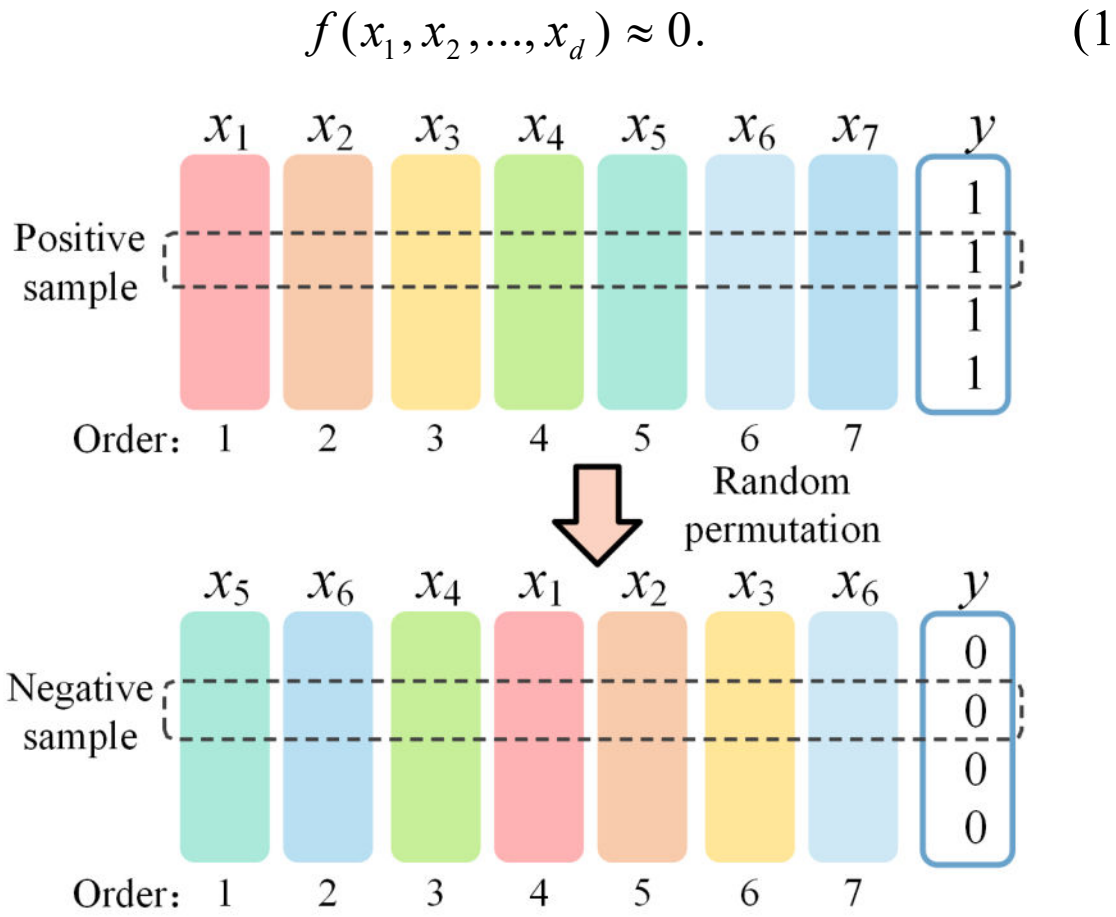

Fig. 4. The process of constructing positive and negative samples.

To solve this problem, KAN uses a contrastive learning approach to transform the unsupervised learning problem into a supervised learning problem taking all variables as the model inputs. For this purpose, KAN defines positive samples as vectors that satisfy domain knowledges or physical rules, which are labeled by 1. Meanwhile, KAN defines negative samples as vectors that are randomly permutated from positive samples, which violate domain knowledges and are labeled by 0. As an example, the sample creation process is illustrated in Fig. 4 assuming there are 7 variables, $x_1$, $x_2$,…, $x_7$.

The KAN model is trained based on the created samples, with the activation function in the last layer being set to Gaussian function. As an example, Fig. 5 shows a KAN network with structure [7,1,1]. After the model training, the magnitudes of the activation function $\varphi_5$, $\varphi_6$ and $\varphi_7$ diminish, indicating the variables $x_5$, $x_6$ and $x_7$ are less important. Such a conclusion could be helpful to feature selection and identifying underlying relationship among variables.

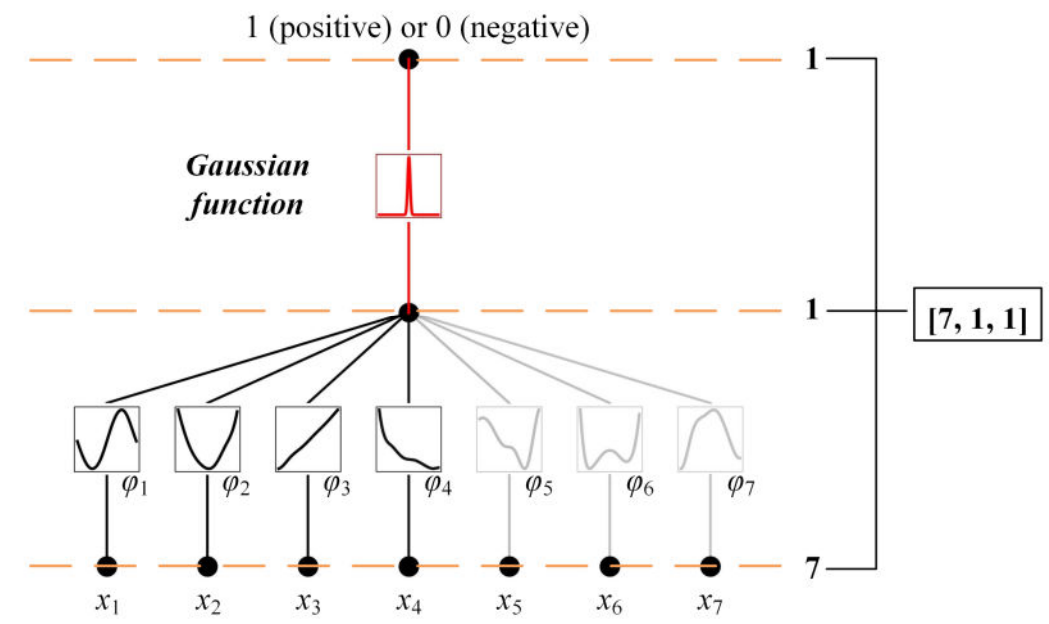

Fig. 5. Example of KAN with structure [7,1,1] in unsupervised learning scenario.

## *E. Improved Symbolification of KAN*

During the symbolification process, the vanilla KAN model calculates $R^2$ between basis functions and the learned activation functions, which is a time-consuming traversal process and can hardly handle the nonlinearity of the complex energy systems [24]. To address this issue, we design a function classifier based on self-attention layers and Mixture of Experts (MoE) mechanisms to replace the $R^2$ calculation for more efficient and accurate symbolification [13], [25].

Self-attention enhances sequence modeling capability by dynamically allocating differential weights to input elements, enabling models to autonomously prioritize salient information. An attention can be formally characterized as a function that maps a query and a collection of key-value pairs to a contextually conditioned output vector through dynamic weight assignment. This operation is typically formulated as:

$$Attention(\mathbf{Q}, \mathbf{K}, \mathbf{V}) = softmax(\frac{\mathbf{QK}^{\mathrm{T}}}{\sqrt{d_k}})\mathbf{V}, \tag{16}$$

where **Q**, **K**, and **V** are learnable parameters matrixes, $d_k$ is the dimension of queries and keys.

MoE has emerged as the predominant paradigm owing to dynamic routing mechanisms that can facilitate parameter-sharing among expert networks. A typical MoE layer comprises a collection of expert networks alongside a gating network. Given that the function matching process necessitates the consideration of all relevant characteristics, a dense MoE layer is selected to achieve a comprehensive outcome. In a dense MoE layer, each expert network is utilized to produce its respective score. The gating network then assigns weights to these scores to aggregate to the final output. MoE can be formulated as

$$\alpha(\mathbf{x}) = Softmax(g(\mathbf{x})), \tag{17}$$

$$y = \sum_{i=1}^{N} \alpha_i(\mathbf{x}) \cdot f_i(\mathbf{x}), \tag{18}$$

where $g(\mathbf{x})$ is the gating network, $\alpha(\mathbf{x})$ is the weight of every export, and $f_i(\mathbf{x})$ is the $i$-th export network.

As illustrated in Fig. 6, the proposed function classifier includes three MoEs, with the feed-forward networks composed by MLPs. Such a specially-designed function classifier can identify the most appropriate basis function to approximate the learned activation function. Such classifier is efficient because it transforms the function selection process into a forward-computation process of neural networks, making the KAN inferencing time independent from the function library size. As such, the KAN computation efficiency can be maintained even as the library size grows.

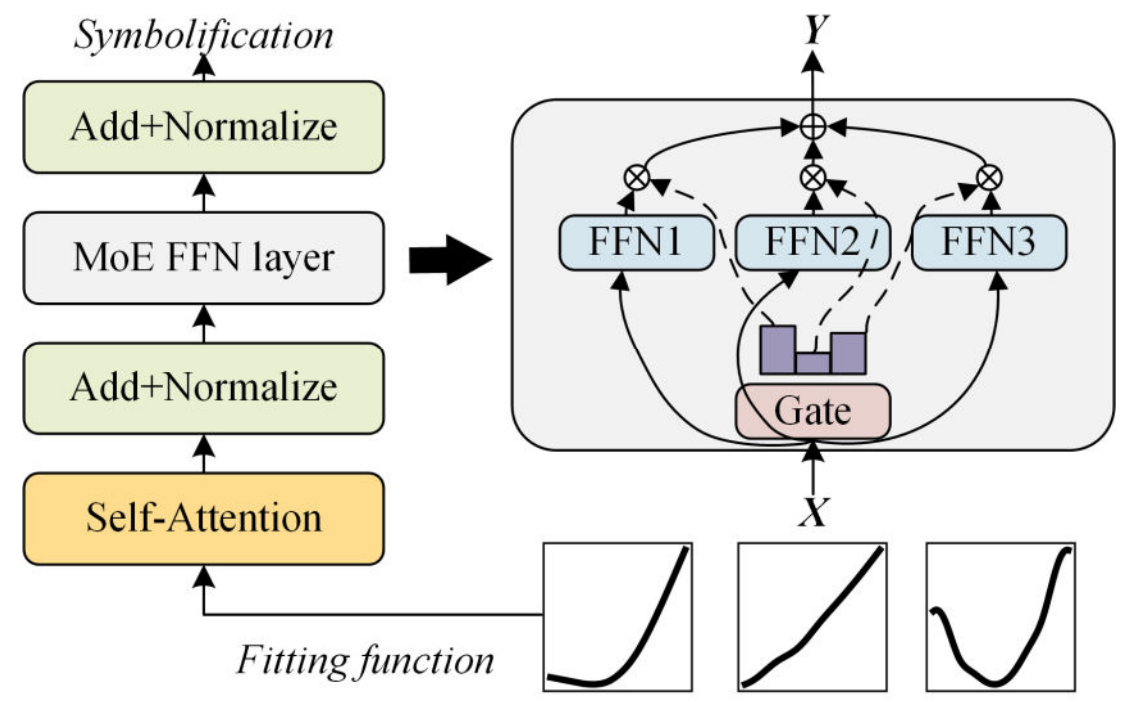

Fig. 6. The function classifier based on attention and MoE.

## III. CASE STUDY

In this section, we evaluate the KAN performance based on 3 test cases: 1) single-phase shift control of a dual active bridge, 2) state of charge estimation of a lithium-ion battery, 3) photovoltaic conversion process of a solar system. Particularly, we focus on the accuracy, interpretability, robustness and generalization of KAN.

### *A. Single-Phase Shift Control of Dual Active Bridge*

Dual Active Bridge (DAB) converter has been widely applied due to its high efficiency and rapid regulation capability [26]. The structure of DAB is illustrated in Fig. 7. The parameters in the DAB are defined as follows: $f$ is the switching frequency, $V_{in}$ is the input voltage, $V_{out}$ is the output voltage, $L$ is the inductance, $n$ is the transformer ratio, $P$ is the transmission power, $D$ is the phase shift ratio between the two H-bridges and is the ratio of the phase shift angle to $2\pi$.

Among the control strategies of DAB, Single Phase Shift (SPS) control is the most basic and widely used control method with fewer control variables [27]. Through the modulation of $D$, SPS can regulate the output voltage and the power flow of DAB. In this test case, we use KAN to construct an explicit mapping function from the phase shift ratio $D$ to the output voltage $V_{out}$, while the other variables are set constant. The simulation system is built by MATLAB/Simulink. For control efficiency, the range of $D$ is limited to [0.3, 0.7].

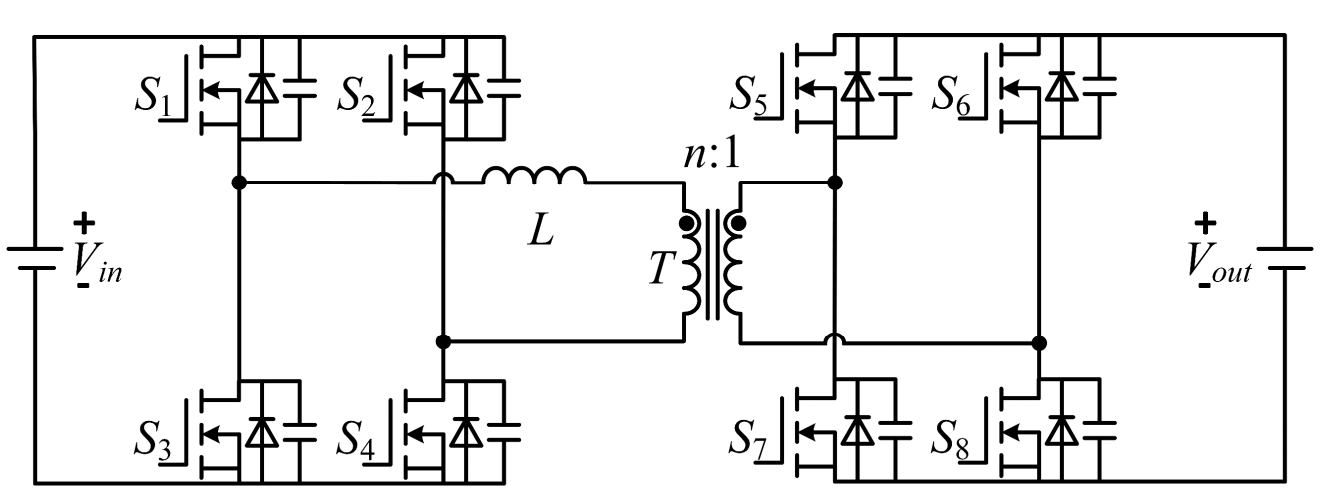

Fig.7. Topology of a DAB converter.

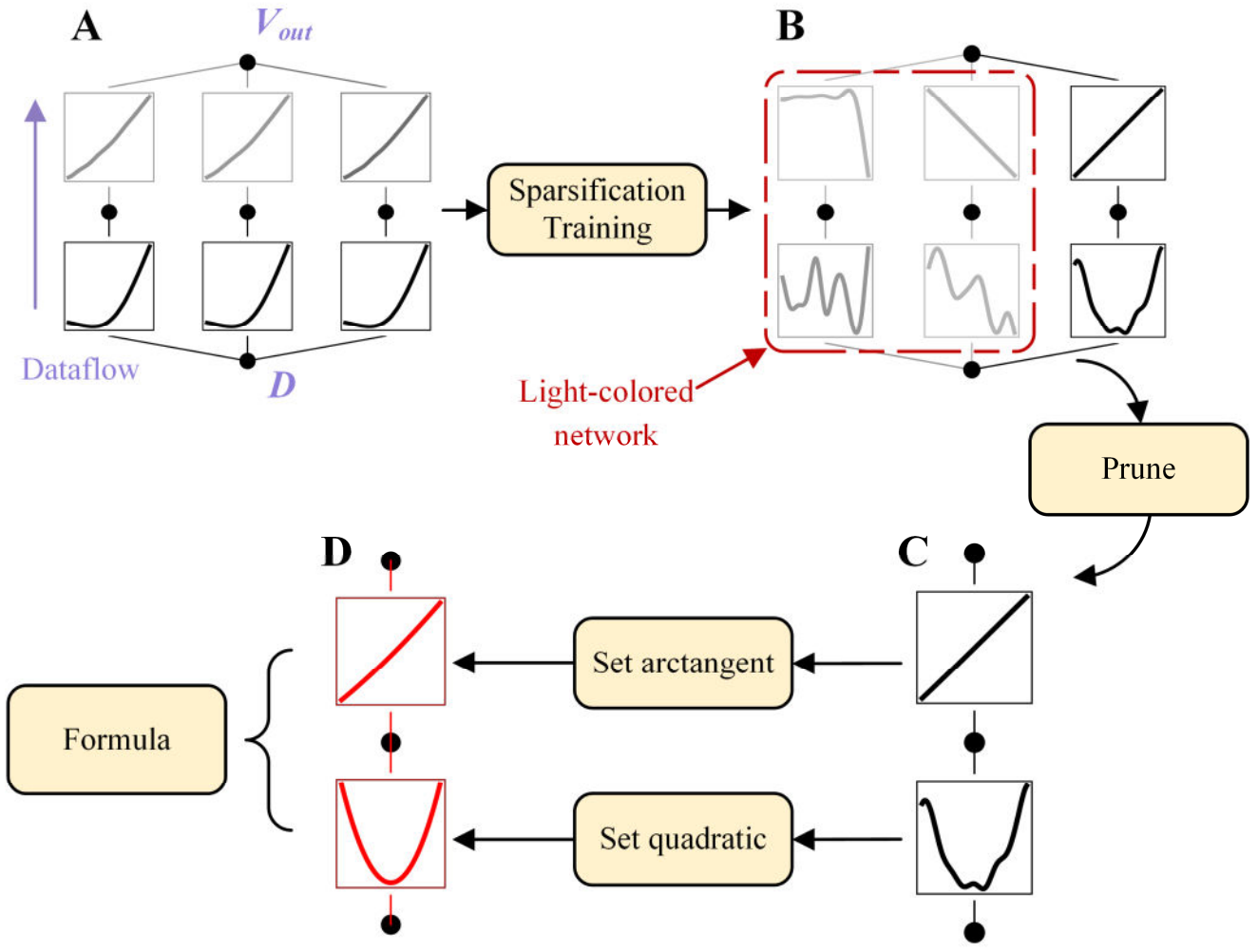

Fig. 8. The training process of KAN in SPS control of DAB. (A) KAN initialization with [1,3,1]. (B) The network after sparsification training. (C) The network [1,1,1] after pruning the light-colored network. (D) The final network after symbolification.

Based on the simulation system, 50000 samples of data pair [$V_{out}$, $D$] are created, of which 80% samples are randomly selected to train the KAN network and the rest 20% are the testing set. The training process of KAN is summarized in Fig. 8. The KAN network is initialized with a structure [1,3,1], and is pruned to [1,1,1] after sparsification. In this case, the symbolic process utilizes a predefined function library consisting of 22 common mathematical functions, such as trigonometric and exponential functions. Then the activation functions are approximated by arctangent function and quadratic function respectively during the symbolification process by searching from the function library. Finally, after fine-tuning the parameters, the final KAN function is

$$V_{out} = 9.04 arctan(8.7(0.5-D)^2 - 1.21) + 15.96. \quad (19)$$

#### *1) Interpretability*

According to the physical process of DAB [28], [29], the ground-truth relationship between $D$ and $V_{out}$ can be explicitly modeled by

$$V_{out} = \frac{2LPf}{nV_{in}D(1-D)} \quad (20)$$

Equation (19) can be re-written as:

$$V_{out} = -9.04 arctan(8.7D(1-D) + 0.965) + 15.96. \quad (21)$$

By comparing equation (20) and (21), we can see that the product term $D$(1-$D$) is successfully learned by KAN. Secondly, although the reverse proportional relationship between $D$(1-$D$) and $V_{out}$ is not captured, KAN instead introduce arctangent function with negative coefficient as the external function, leading to similar function monotonicity when $D \in (0,1)$. This means KAN has the capability to learn the physical laws of an electrical energy system based on field measurement data without any prior knowledge. The results are given by explicit and concise symbolic equations, which is a "glass-box" and has much higher interpretability compared with traditional "closed-box" neural networks.

#### *2) Accuracy and Robustness*

We further evaluate the accuracy and robustness of KAN and compare its performance with MLP based on the testing set. Three metrics are calculated to evaluate the model accuracy: Mean Absolute Error (MAE), Root Mean Squared Error (RMSE) and bias.

$$MAE = \frac{1}{n}\sum_{i=1}^{n} |y_i - \hat{y}_i|, \quad (22)$$

$$RMSE = \sqrt{\frac{1}{n}\sum_{i=1}^{n}(y_i - \hat{y}_i)^2}, \quad (23)$$

$$bias = \frac{1}{n}\sum_{i=1}^{n}\frac{y_i - \hat{y}_i}{y_i}, \quad (24)$$

where $y_i$ is the ground truth and $\hat{y}_i$ is the predicted data. MAE evaluates the cumulative prediction error, RMSE focuses on the point-to-point prediction error, and bias reflects the presence of a systematic discrepancy between real data and model predicted data. In addition, to evaluate the model robustness, we add random noises with magnitudes less than 10% to $D$ to see the model performance under perturbation. Results are summarized in Table I.

In terms of RMSE and MAE, from Table I, we can see that KAN shows superior accuracy in both the training and testing set in the noise-free scenario, compared with MLP. However, the performance of KAN degrades on the noised data set. This is because KAN is a compact model with only a few parameters and limited complexity. As a result, KAN is more sensitive to the data quality and has higher risk to be biased by abnormal samples. On the contrary, MLP has large number of parameters and much higher complexity, showing higher robustness and greater resistance to noises. This inspires us to be more careful on data preprocessing when training the KAN model.

TABLE I
MODEL PERFORMANCES ON THE TRAINING AND TEST SET (DAB)

| Model | Metric | Training set | Test set | Training set(noise) | Test set(noise) |
|---|---|---|---|---|---|
| KAN | MAE | 0.0041 | 0.0041 | 0.4149 | 0.4206 |
| | RMSE | 0.0042 | 0.0042 | 0.6077 | 0.6161 |
| | bias | -0.0005 | -0.0004 | -0.0050 | -0.0038 |
| MLP | MAE | 0.0193 | 0.0189 | 0.3304 | 0.3348 |
| | RMSE | 0.0333 | 0.0327 | 0.4339 | 0.4376 |
| | bias | -0.0001 | -0.0002 | -0.0095 | -0.0091 |

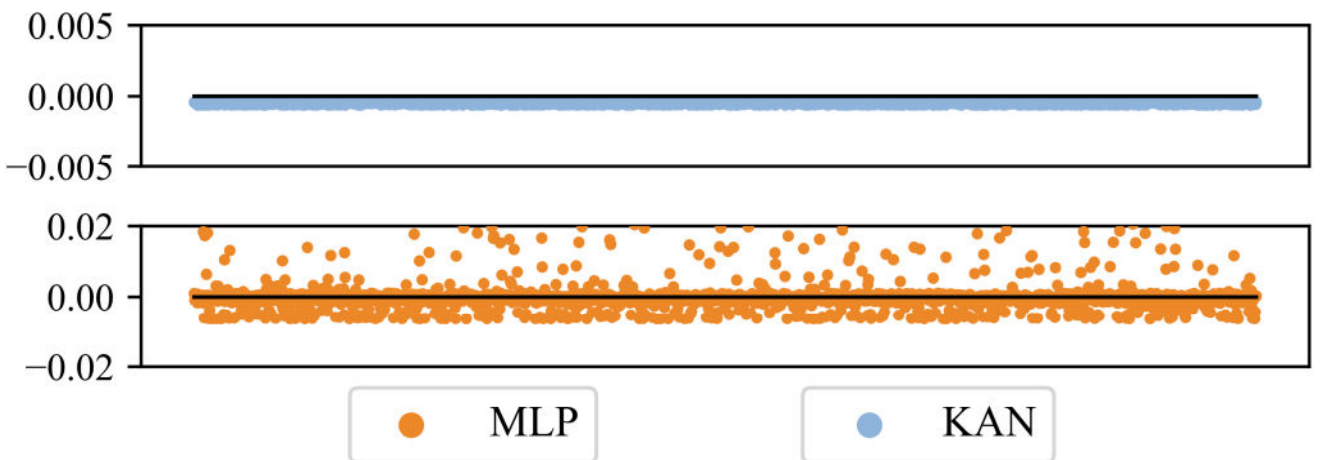

Fig. 9. Point-by-point biases of KAN and MLP. The black horizontal dashed line represents the zero bias.

Concerning bias, the opposite trend is observed. The MLP exhibits a lower systematic discrepancy and is more responsive to fluctuations compared to KAN. Indeed, as illustrated in Fig. 9, the fitting results of KAN are generally slightly below the ground truth, whereas the MLP's predictions acutely fluctuate around the actual values. Overall, KAN derives a mathematical formula through fitting, which, although predicts with systematic bias, achieves higher accuracy than the MLP. In contrast, as a high-parameter closed-box model, the MLP yields fitting results devoid of systematic bias; however, its errors vary significantly, indicating larger deviations.

*3) Generalization*

Generalization is another key indicator to evaluate whether the model learns the true pattern that can be applicable outside the training data set. In this test case, we train both MLP and KAN based on the samples that $D \in [0.3, 0.7]$. To evaluate the model generalization capability, we test the model performance based on samples that $D \in (0.2, 0.3] \cup [0.7, 0.8)$. Results are shown in Fig. 10. We can see that the fitting accuracy of KAN is still satisfying when the testing samples falls outside the training interval, while the fitting error of MLP increases significantly. This is because KAN learns the true laws of the physical process that can be generalized to a wider range. On the contrary, MLP tends to be overfitting and has weak generalization capability, due to its large amount of parameters and complex model structures. Meanwhile, due to the quadratic function, KAN learns a symmetry relationship, which MLPs are unable to capture.

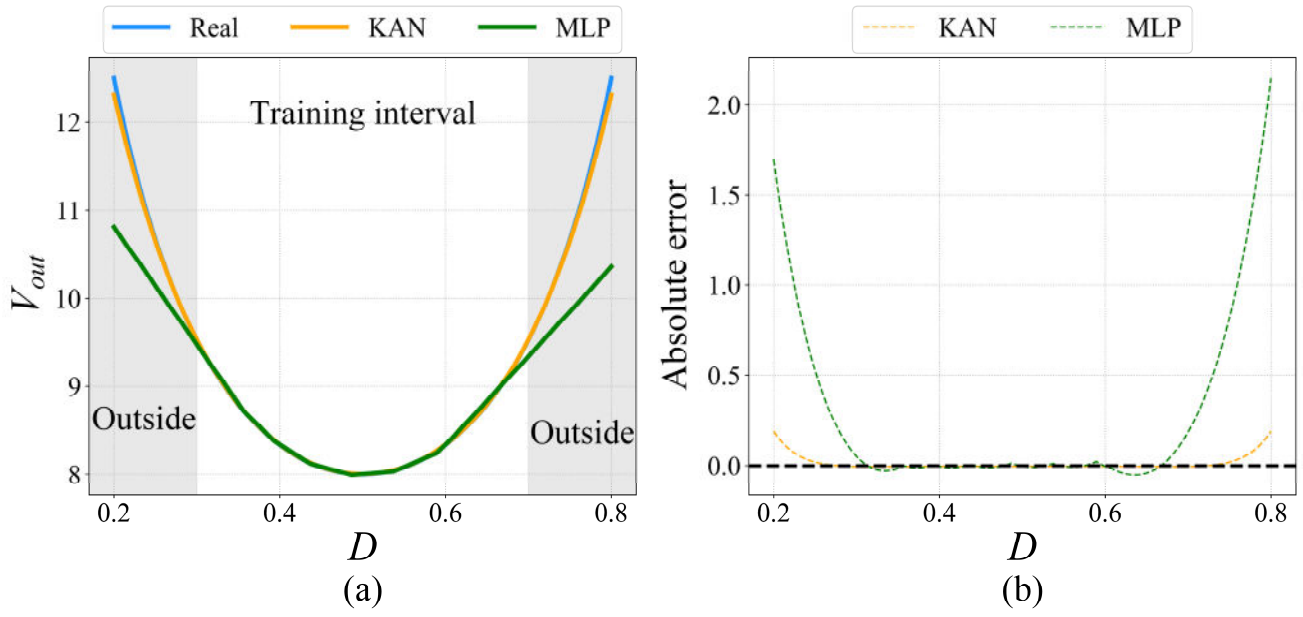

Fig. 10. Fitting results of KAN and MLP. (a)The overall trends, (b)the absolute errors of KAN and MLP.

*4) Sensitivity analysis of model configuration*

Configuring hyper-parameters for a neural network is generally considered empirical and may significantly influence the model performance. In this section, we further evaluate the KAN model performance with different structures to see whether KAN is sensitive to the model configuration. According to equation (20), $V_{out}$ is dependent on both $D$ and (1-$D$). Here we set $D$ and (1-$D$) as two separate inputs of the KAN model with an initialized network structure [2,5,1] to test the model performance. After similar training process, the KAN model can be expressed as

$$V_{out} = 27.46 - 22.33 sigmoid(2.37(0.46 - (1-D))^2 - 15.04(0.51-D)^2 + 1.91), \tag{25}$$

which can be re-written as

$$V_{out} = 27.46 - 22.33 sigmoid(-1.34 + 12.78(D(1-0.99D))). \tag{26}$$

Based on (26), we can see that the key product term $D$(1-$D$) is also captured by KAN even under different model input and structure configuration. The fitting results of the two different KAN models are compared in Fig. 11. Both two KAN models show satisfying accuracy. This means KAN has tolerance for model configuration, which is friendly to engineers in practice.

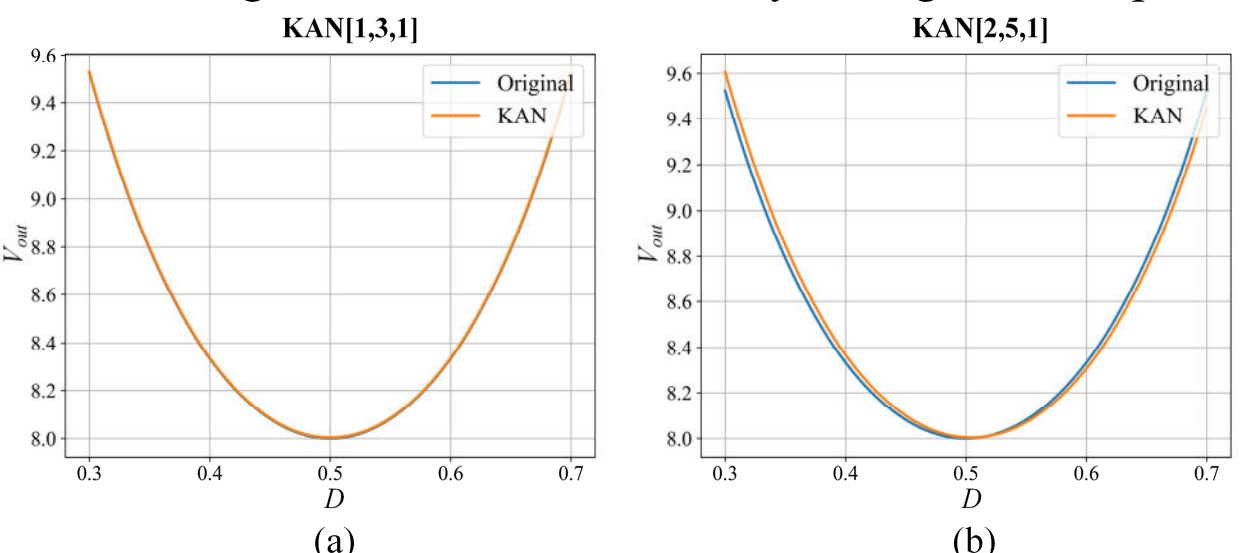

Fig. 11. KAN fitting results under different configurations. (a) KAN model with initialized structure [1,3,1] and D as the single input. (b) KAN model with initialized structure [2,5,1] and D, (1-D) as double inputs.

*5) Neural network function classifier*

The initial KAN selects basic functions based on their $R^2$ correlation scores. Table II illustrates the $R^2$ correlation scores between various basis functions and the fitted functions. Note that the top six $R^2$ correlation scores are very close, particularly for those in the second position. Such proximity suggests that selecting the most suitable basis function solely based on $R^2$ correlation scores may be less sensitive.

To address this issue, we trained an additional KAN model with the configuration [1,3,1], incorporating the proposed neural network–based function classifier. In accordance with the function library, the classifier is set up to select one of the 22 common functions in the function library. Results are shown as equation (27). It can be seen that the neural network function classifier identified the inverse proportional function as the most appropriate basis function for the KAN model.

TABLE II
THE TOP SIX $R^2$ CORRELATION SCORES

| Position | (0, 0, 0) | | (1, 0, 0) | |
|---|---|---|---|---|
| Rank | Function | $R^2$ | Function | $R^2$ |
| 1 | $x^2$ | 0.9953 | arctan | 0.9975 |
| 2 | cos | 0.9950 | tanh | 0.9974 |
| 3 | gaussian | 0.9948 | sigmoid | 0.9974 |
| 4 | sin | 0.9945 | gaussian | 0.9972 |
| 5 | abs | 0.9299 | sin | 0.9953 |
| 6 | $x^4$ | 0.9211 | 1/x | 0.9953 |

$$V_{out} = \frac{2.38}{(0.30 - 1.19*(0.5-D)^2)} = \frac{2}{0.002 + D(1-D)}. \quad (27)$$

By comparing equation (20), (21) and (27), equation (27) is considered more interpretable and appropriate to describe DAB, because equation (27) is more similar to equation (20). Regarding the computation efficiency, the symbolification process of the vanilla KAN takes 19.11s, while the proposed neural network function classifier requires only 2.58s. implementing the classifier significantly accelerates the symbolification process.

## B. State of Charge Estimation of Lithium-ion Battery

Dara-driven approaches have been widely used for the state of charge (SOC) estimation of lithium-ion batteries [30], [31], [32], of which the key is to establish the mapping from field measurements to SOC values. In this section, we implement KAN to achieve SOC estimation for a lithium-ion battery system to further evaluate its effectiveness.

The data set is based on a Panasonic 18650PF Li-ion battery under discharging condition [33]. We set *SOC* as the dependent variable, and the battery terminal voltage *U* and discharging current *I* as the independent variables. It should be noted that the discharge current is set to negative values. The data set is split to 80% and 20% for KAN model training and testing, respectively.

The KAN network is initialized with a structure of [2,5,1], and the final expression after training is

$$SOC = 0.26\left| \begin{matrix} 6.15 sigmoid(2.78U - 10.01) \\ + 2.23 sigmoid(-0.54I - 0.39) - 3.05 \end{matrix} \right| + 0.3. \quad (28)$$

### 1) Interpretability

Based on equation (28), we can see that the lower bound of the SOC estimation results is 0.3, which is the same as the minimum SOC value in the data set. Besides, according to (29), the SOC is positively correlated with the terminal voltage *U*, and is negatively correlated with the discharging current *I*. The magnitude of the coefficient of *U* is larger than that of *I*, indicating that SOC is more sensitive to the change of terminal voltage. All the above conclusions that obtained from equation (28) can be correlated with the physical characteristics of the battery system, showing again that KAN has significant advantages with respect to interpretability. The plots of KAN fitting results are shown in Fig. 12.

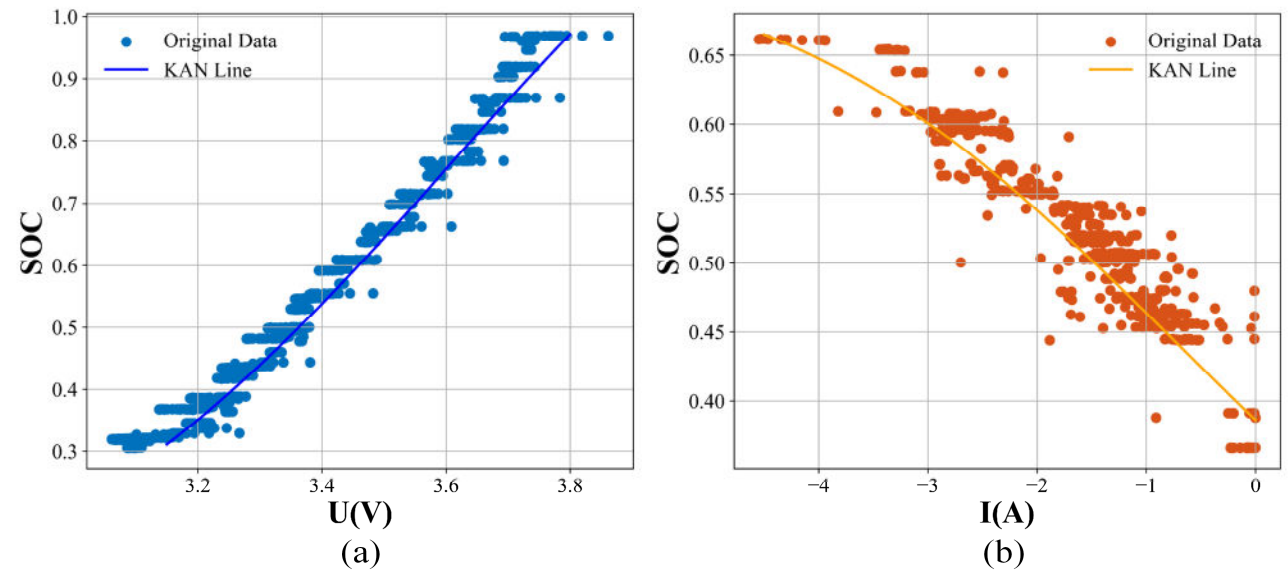

Fig. 12. Plots of KAN fitting results. (a) SOC variation with U when I is fixed at -3A. (b) SOC variation with I when U is fixed at 3.45V.

### 2) Accuracy and Robustness:

Similar to case 1, we evaluate the KAN model accuracy by calculating three error metrics, and evaluate the model robustness by adding noises to the original data sets. MLP is also used as the benchmark method. Specially, for lithium-ion battery, some physical-informed methods have been proposed[34], [35].

TABLE III
MODEL PERFORMANCES ON THE TRAINING AND TEST SET (BATTERY)

| Model | Metric | Training set | Test set | Training set(noise) | Test set(noise) |
|---|---|---|---|---|---|
| KAN | MAE | 0.0132 | 0.0134 | 0.0428 | 0.0429 |
| | RMSE | 0.0169 | 0.0174 | 0.0539 | 0.0540 |
| | bias | -0.0123 | -0.0121 | -0.0174 | -0.0145 |
| MLP | MAE | 0.0127 | 0.0131 | 0.0443 | 0.0450 |
| | RMSE | 0.0170 | 0.0177 | 0.0557 | 0.0561 |
| | bias | 0.0128 | 0.0128 | 0.0122 | 0.0134 |
| PINN | MAE | 0.0302 | 0.0282 | 0.0511 | 0.0499 |
| | RMSE | 0.0354 | 0.0339 | 0.0575 | 0.0559 |
| | bias | 0.0143 | 0.0116 | 0.0213 | 0.0205 |

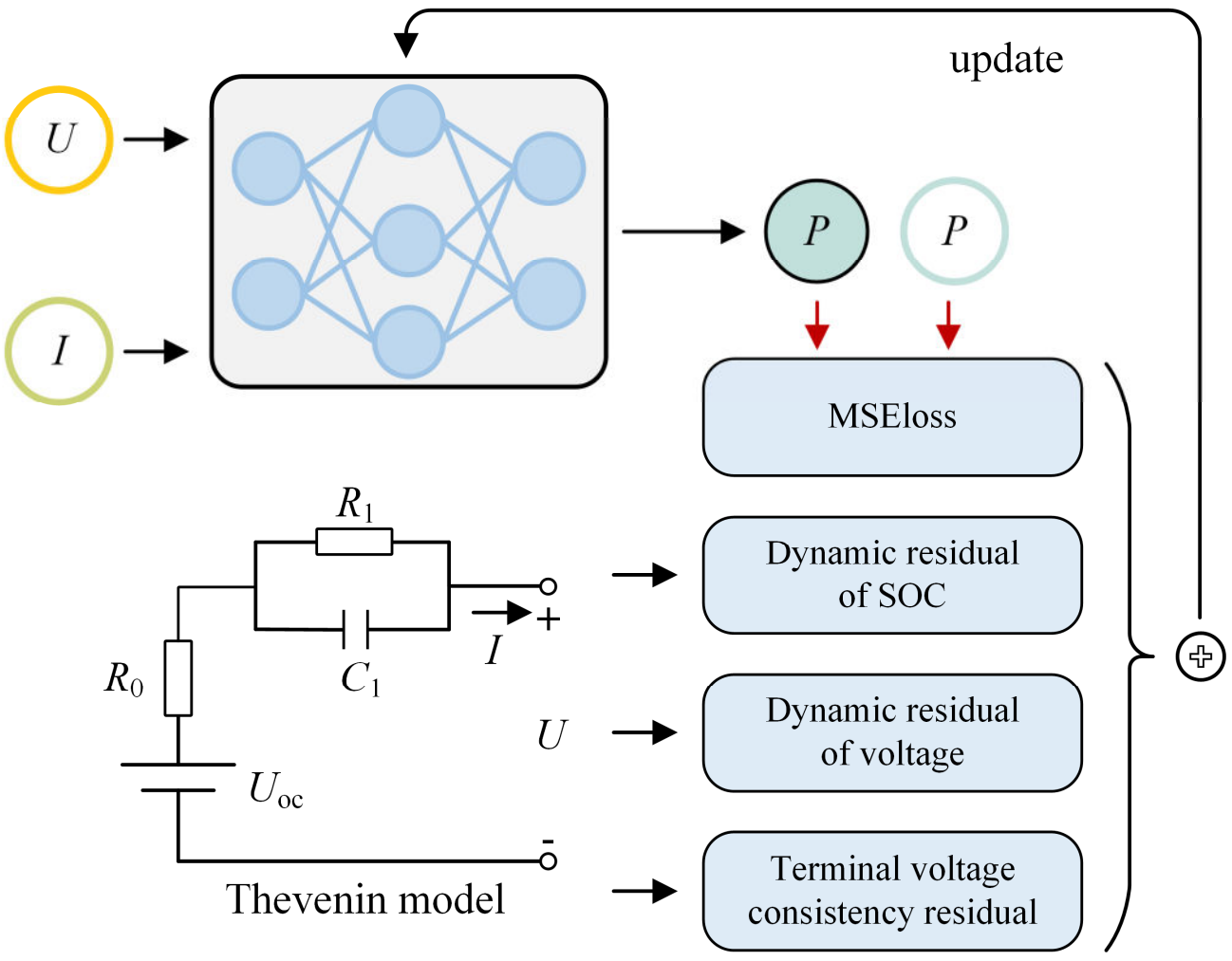

Fig. 13. The PINN of lithium-ion battery.

As shown in Fig. 13, we design a simple PINN by incorporating the physical equation as a regularization term based on Thevenin model [36], [37], used as an additional benchmark. Results are summarized in Table III. We can see that KAN shows better performance than MLP and PINN in both accuracy and robustness.

## C. Photovoltaic Process of Solar Systems

Photovoltaic process of solar systems is a complex physical process influenced by multiple factors such as solar irradiance, temperature, wind speed, etc. In this section, we first implement unsupervised-learning KAN to identify the key factors that dominate the power output of the solar system. Then we further implement supervised-learning KAN to establish the mapping from the identified factors to the power output to simulate the photovoltaic process. The data set we use is the DKASC dataset [38], of which 80% for model training and 20% for model testing. We implement min-max normalization to preprocess the data set to align the scales of different variables.

### 1) Unsupervised-learning KAN for feature selection

We initialize a KAN network with structure [7,1,1] to mine the underlying correlations between solar power output and the other 6 meteorological factors. After training, the KAN network shrinks to [4,1,1] shown as Fig. 14, indicating that wind speed, temperature, radiation are the most influential factors to the solar power output.

As a comparison, we also calculate another 3 correlation indicators between the solar power output and each of the 6 meteorological factors: Pearson correlation coefficient, Spearman rank correlation coefficient and Kendall rank correlation coefficient [39], [40], [41]. Results are summarized in Table IV. We can see that the feature identification results of KAN are in accordance with the correlation coefficients, indicating that KAN is able to identify the dependency among variables without being given any prior knowledge.

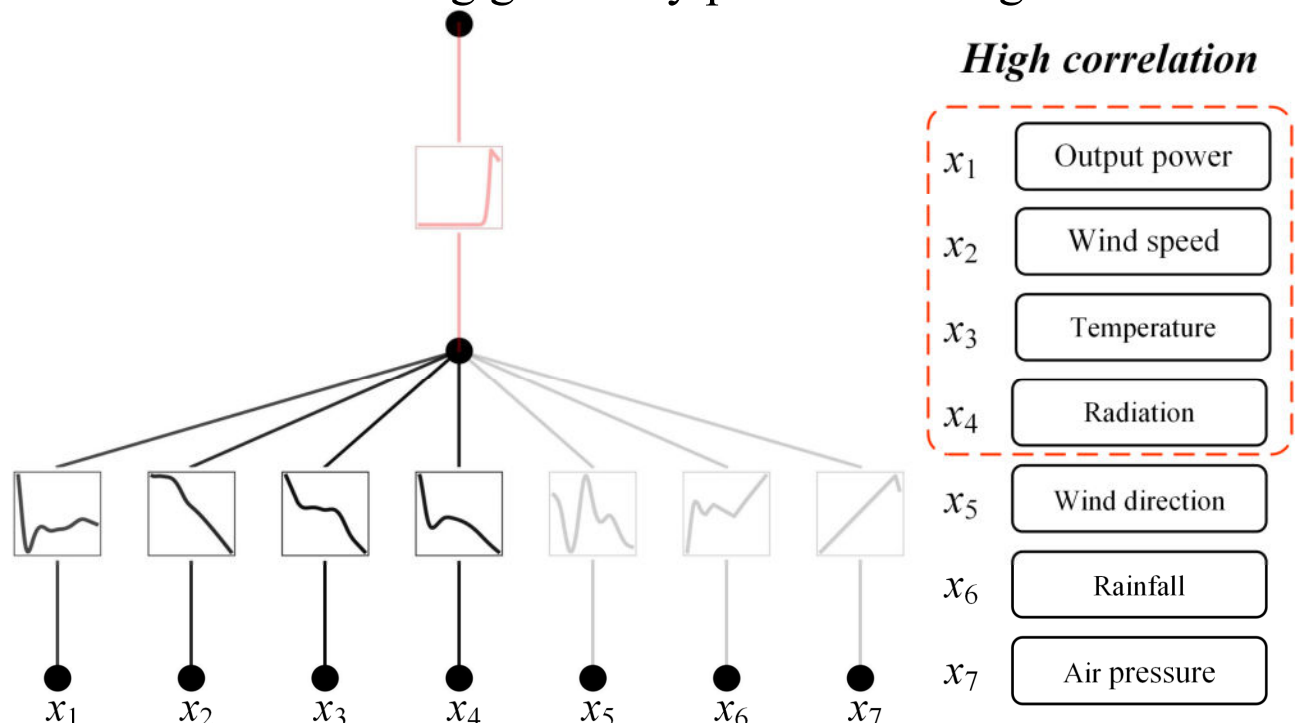

Fig. 14. Functional dependency between solar power output and meteorological factors identified by unsupervised-learning KAN.

TABLE IV
CORRELATION BETWEEN METEOROLOGICAL VARIABLES AND POWER OUTPUT

| Variables | Pearson | Spearman | Kendall |
|---|---|---|---|
| Wind speed | 0.4828 | 0.4615 | 0.3118 |
| Temperature | 0.4037 | 0.5498 | 0.3997 |
| Radiation | 0.9283 | 0.7907 | 0.6226 |
| Wind direction | -0.1305 | -0.2065 | -0.1366 |
| Rainfall | -0.0339 | -0.0368 | -0.0301 |
| Air pressure | -0.0233 | -0.0122 | -0.0103 |

### 2) Interpretability

Based on the feature selection results, we further train a supervised-learning KAN model with the temperature $T$, radiation $R$ and wind speed $v$ as the model inputs and the power output $P_{out}$ as the model output to simulate the photovoltaic process. The KAN network is initialized by [3,7,1] and shrinks to [3,1,1] after training. The final expression is

$$\begin{aligned} P_{out} &= 0.69 + 0.48 arctan(0.04 sin(10.18v - 5.08) \\ &+0.93 tan(2.62R - 1.46) - 0.19 arctan(8.73T - 4.6) \\ &+0.24). \end{aligned} \quad (29)$$

From equation (29) we have the following observations: 1) The range of $P_{out}$ is strictly constrained by the external arctangent function, which is in accordance with the normalized power output of the solar system. 2) $P_{out}$ is positively correlated with radiation $R$, and is negatively correlated with temperature $T$. Note that the wind speed $v$ has a positive impact on the power output when $v$ is under a certain threshold. However, when $v$ is too large, the impact may turn to be negative due to possible damages to the solar system.

TABLE V
AVERAGE SENSITIVITY OF INDEPENDENT VARIABLES

| **Variables** | Wind speed $v$ | Radiation $R$ | Temperature $T$ |
|---|---|---|---|
| **Sensitivity** | 0.0707 | 1.2582 | 0.1468 |

We further conduct sensitivity analysis to quantify the influence of $R$, $T$, $v$ to $P_{out}$ [42]. Results are shown in Table V. We can see that radiation is the most influential factor, followed by temperature and wind speed. To summarize, all the above conclusions obtained from the KAN modeling results in equation (29) is in accordance with the domain knowledge of a PV system, which again demonstrates the interpretability of KAN.

### 3) Accuracy and Robustness

As a benchmark, we design a PINN based on thermal and photoelectric efficiency models, which comprehensively considers the influence of temperature, wind speed and irradiance [43], [44]. As shown in Fig. 15, the physical and informed models are jointly trained by aggregating their results, calculating the combined loss, and simultaneously updating their parameters.

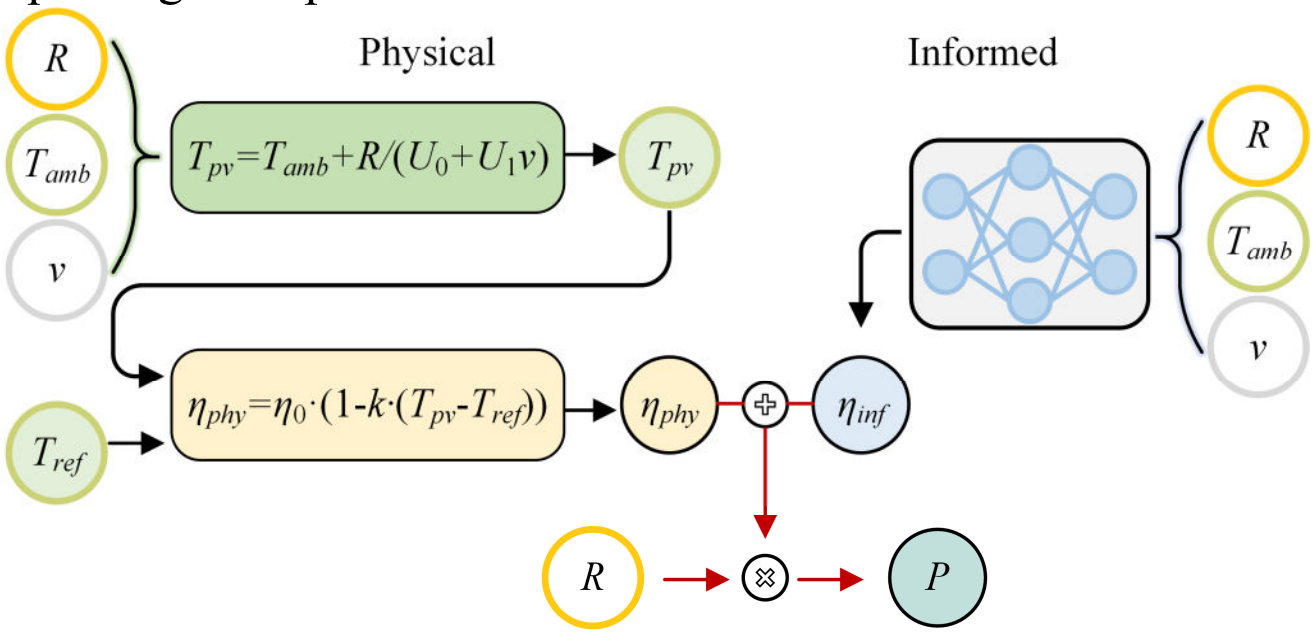

Fig. 15. The PINN of solar system.

TABLE VI
MODEL PERFORMANCES ON THE TRAINING AND TEST SET (PV)

| Model | Metric | Training set | Test set | Training set(noise) | Test set(noise) |
|---|---|---|---|---|---|
| **KAN** | **MAE** | 0.0380 | 0.0386 | 0.0558 | 0.0555 |
| | **RMSE** | 0.0903 | 0.0918 | 0.1298 | 0.1256 |
| | **bias** | -0.0086 | -0.0089 | -0.0095 | -0.0096 |
| **MLP** | **MAE** | 0.0397 | 0.0403 | 0.0523 | 0.0526 |
| | **RMSE** | 0.0909 | 0.0904 | 0.0944 | 0.0953 |
| | **bias** | -0.0063 | -0.0069 | -0.0065 | -0.0070 |
| **PINN** | **MAE** | 0.0354 | 0.0349 | 0.0612 | 0.0615 |
| | **RMSE** | 0.0813 | 0.0796 | 0.1053 | 0.1056 |
| | **bias** | -0.0024 | -0.0024 | -0.0176 | -0.0181 |

The accuracy metrics are calculated and summarized in Table VI. We can see that KAN has comparable accuracy with MLP and PINN while maintaining a significant advantage on model interpretability.

## IV. CONCLUSION

In this paper, we introduce KAN to achieve end-to-end “glass-box” modeling for electrical energy systems. The interpretability of KAN mainly comes from two aspects. *First*, the sparse training process reduces KAN from a complex initialization network to a much-simplified network with only a few edges remained, which enables the interpretation of the modeling results. *Second*, the symbolification process replaces the learned activation functions on the remained edges with common basis functions, which have clearer physical meanings and further improves the interpretability. For improve efficiency and interpretability, a function classifier consisted of attention and MoE is designed to take place of repetitive $R^2$ scores in KAN framework. As a result, KAN can express the physical process with concise and explicit mathematical formulas, revolutionarily improving the interpretability compared with traditional deep learning models. Simulation results based on three different electrical energy systems demonstrate that, compared with MLP and PINN, KAN has superior performance in the aspects of interpretability and generalization ability while keeping comparable accuracy and robustness.

Future work may focus on further evaluating the performance of KAN in a wider range, as well as improving the training efficiency and stability.